\documentclass[aip,
nofootinbib,
rsi,%
reprint,
twocolumn,
longbibliography,
author-numerical%
]{revtex4-2}

\usepackage[utf8]{inputenc}
\usepackage[T2A]{fontenc}
\usepackage{graphicx}
\usepackage{textcomp}

\usepackage{amsfonts}
\usepackage{amssymb}
\usepackage{amsmath}
\usepackage{color}

\usepackage{changes}

\newcommand{\pdr}[2]{\dfrac{\partial {#1}}{\partial #2}}

\newcommand{\pddr}[2]{\dfrac{\partial^2{#1}}{\partial #2^2}}

\newcommand{\pdra}[2]{{\partial    #1}/{\partial #2}}

\newcommand{\tx}{\tilde{x}}

\newcommand{\tz}{\tilde{z}}

\newcommand{\tit}{\tilde{t}}

\newcommand{\tom}{\widetilde{\omega}}

\newcommand{\tj}{\tilde{j}}

\newcommand{\tc}{\tilde{c}}

\newcommand{\teta}{\tilde{\eta}}

\newcommand{\tJ}{\widetilde{J}}

\newcommand{\tZ}{\widetilde{Z}}
\newcommand{\tD}{\widetilde{D}}
\newcommand{\tN}{\widetilde{N}}
\newcommand{\tl}{\tilde{l}}

\newcommand{\expn}{{{\rm e}^{\teta^0}}}

\newcommand{\cref}{c_{ref}}

\newcommand{\jlim}{j_{\lim}}

\newcommand{\Cdlast}{C_{dl}^*}
\newcommand{\Cdl}{C_{dl}}

\newcommand{\tjlim}{\tilde{j}_{\lim}}

\newcommand{\tCdl}{\widetilde{C}_{dl}}

\newcommand{\lam}{\lambda}

\newcommand{\veps}{\varepsilon}

\newcommand{\sion}{\sigma_p}
\newcommand{\tsion}{\tilde{\sigma}_p}
\newcommand{\sionast}{\sigma_p^*}

\newcommand{\Dox}{D_{ox}}
\newcommand{\tDox}{\tD_{ox}}

\newcommand{\matr}[1]{\mathbf{#1}}
\newcommand{\matU}{\matr{U}}
\newcommand{\matM}{\matr{M}}
\newcommand{\vecv}{\matr{v}}

\newcommand{\lcat}{l_t}

\newcommand{\ri}{{\rm i}}

\newcommand{\eeo}{{\rm e}^{\teta^0}}

\newcommand{\sqa}{\ri\tom/\left(\veps^2\tD_b\right)}

\newcommand{\lexp}[1]{\exp\left(#1\right)}
\newcommand{\lnl}[1]{\ln \left(#1\right)}

\newcommand{\lcosh}[1]{\cosh\left(#1\right)}
\newcommand{\lsinh}[1]{\sinh\left(#1\right)}

\newcommand{\RMaple}{Maple$^{\circledR}${ }}

\newcommand{\etal}{et al.{ }}

\begin{document}

\sf

\title{A fast physics-based matrix model for the impedance of a PEM fuel cell:
       Incorporating functionally graded catalyst layer and channel impedances}

\author{Andrei Kulikovsky}
\email{A.Kulikovsky@fz-juelich.de}

\affiliation{Forschungszentrum J\"ulich GmbH             \\
    Theory and Computation of Energy Materials (IET-3)   \\
    Institute of Energy and Climate Research,            \\
    D-52425 J\"ulich, Germany
}

\begin{abstract}
We extend a recent physics-based matrix model for calculating PEM fuel cell impedance
(doi:10.1149/2754-2734/ad6ce8) to cases
of low air flow stoichiometry and functionally graded  cathode catalyst layers (CCLs).
We demonstrate that the matrix model produces
accurate spectra and is almost three orders of magnitude faster than a model based
on the standard boundary-value problem solver. The physics-based matrix model
can compete with equivalent circuit models for fitting experimental EIS spectra,
particularly those measured from cells with functionally graded CCL.
\end{abstract}


\maketitle

\section{Introduction}

Electrochemical impedance spectroscopy has proved to be an invaluable tool
for the characterization and testing of PEM fuel cells. However, after
decades of research, deciphering impedance spectra remains a challenging task.

The major potential losses in PEMFCs occur in the cathode catalyst layer (CCL).
The distribution of the electrochemical reaction through the CCL depth
is typically non-uniform.
To mitigate the losses associated with this non-uniformity, there is
growing interest in developing functionally graded
catalyst layers \cite{Song_05,Yoon_10,Kulikovsky_12j,Ni_23,Ayoub_24}.

One of the simplest and widely used approaches to understanding
the impedance spectra
is equivalent circuit modeling (ECM) \cite{Page_07,Han_10,Kim_13,Minggao_21}.
However, despite numerous equivalent circuits
suggested for different cell operating conditions \cite{Han_10,Kim_13,Heinzmann_18},
the ECM is not fully reliable\cite{Macdonald_06}. The application
of ECM to catalyst layers with the non-uniform parameters is particularly
questionable.

Over the past two decades, a model-free distribution
of relaxation times (DRT) method for the spectra analysis
has emerged \cite{Saccoccio_14,Tiffee_17,Boukamp_17}.
The DRT was invented by Fuoss and Kirkwood in 1941\cite{Fuoss_41}
and brought to the fuel cell community by Schlichlein \etal\cite{Schlichlein_02}.
The method is based on the formal spectrum expansion into an infinite series
of parallel $RC$-circuit impedances. The resulting DRT spectrum exhibits
a number of peaks, with each peak being associated with one or more processes inside the fuel cell.
However, in PEMFCs, the oxygen transport peaks often overlap with the faradaic
peak, while some other peaks may be numerical artefacts. The attribution
of DRT peaks remains a subject of ongoing research.
In non-uniform porous systems, the DRT transport peaks would become wider
as the local characteristic transport time and the double layer (DL) charging time
 varies along the spatial coordinate.
This would make attributing the DRT peaks an even more challenging task.

An alternative to the ECM and DRT is the physics-based (PB)
impedance modeling (see reviews \cite{Zhang_20,Huang_20}).
The PB modeling of the first type is based on
the direct numerical solution of the transient mass
and charge transport equations inside the cell by applying a small harmonic
perturbation of the current \cite{Shi_08}, or the potential step \cite{Bao_15,Futter_18}.
The impedance is calculated using the Fourier transform
of the applied perturbation and the system response.
The second type of PB modeling uses linearization and Fourier-transform of the
transport equations to derive a system of linear equations for small
perturbation amplitudes in the frequency domain \cite{Kulikovsky_15g,Weber_26}.
Solving this system directly yields the impedance.

The PB models of the second type are much faster for least-squares fitting experimental spectra,
than the models of the first type. However, even the second-type models still require
tens of minutes to hours for fitting a spectrum \cite{Kulikovsky_26b}.
For example, calculating a single spectrum with the advanced PB model \cite{Weber_26} takes 26 sec.
Least-squares fitting algorithms require the calculation of hundreds of spectra to find
optimal fitting parameters.

The impedance spectra of PEMFCs with the functionally graded CCL exhibit
features that can be best described using PB models for non-uniform
CCLs \cite{Gerteisen_15,Kulikovsky_17a}. This requires development of fast models
for the characterization of such cells and stacks.

Below, we develop a fast PB model for fitting impedance spectra.
The model is based on a novel matrix solver for the system of linear equations
for the perturbation amplitudes in the frequency domain.
The first version of the model
suitable for uniform catalyst layers was reported in \cite{Kulikovsky_24f}. Here,
we extend this model to the functionally graded CCLs.
In addition, we incorporate the impedance of oxygen transport
in the channel into the model. The resulting model is an accurate and efficient
tool for fitting the spectra at high cell currents and low air flow
stoichiometries. The model is nearly three orders of magnitude faster
than the PB model based on standard Python BVP solver. This makes the matrix model
suitable for fast spectra fitting in real-time testing systems.
Particularly important field of the model applications is
EIS characterization of cells with functionally graded catalyst layers.

\section{Model}

The CCL electron conductivity is assumed to be high. This assumption is justified
for typical PEM fuel cells.
However, it does not apply to polymer electrolyte electrolysis cells.

\begin{figure}
\begin{center}
    \includegraphics[scale=0.8]{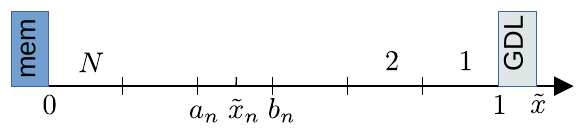}
\end{center}
    \caption{Schematic of the cathode catalyst layer separated into $N$
        numerical cells. The problem \eqref{eq:teta1x2},
        \eqref{eq:tc1x2} is solved in every cell $[a_n, b_n]$.
        Note that the cells are numbered from the GDL/CCL interface
        toward the membrane.
    }
    \label{fig:sketch}
\end{figure}

The core of the model is formed by the transient  conservation equations
for the oxygen concentration and charge in the CCL:
\begin{align}
   & \Cdl\pdr{\eta}{t} - \pdr{}{\tx}\left(\sion\pdr{\eta}{x}\right)
            = - i_* \left(\dfrac{c}{\cref}\right)\lexp{\dfrac{\eta}{b_T}}
            \label{eq:etax} \\
   &\pdr{c}{t} - \pdr{}{\tx}\left(\Dox\pdr{c}{x}\right)
               = - \dfrac{i_*}{4F} \left(\dfrac{c}{\cref}\right)\lexp{\dfrac{\eta}{b_T}}
            \label{eq:cx}
\end{align}
Here
$t$ is time,
$x$ is the coordinate through the cell counted from the membrane (Figure~\ref{fig:sketch}),
$\eta$ is the positive by convention oxygen reduction reaction (ORR) overpotential,
$\Cdl$ is the volumetric double layer capacitance (F~cm$^{-3}$),
$\sion$ is the CCL proton conductivity,
$b_T$ is the ORR Tafel slope,
$i_*$ is the ORR exchange current density,
$c$ is the local oxygen concentration in the CCL,
$\cref$ is the reference oxygen concentration,
$\Dox$ is the effective oxygen diffusion coefficient in the CCL, and
$F$ is the Faraday constant.
The right side of Eqs.\eqref{eq:etax},\eqref{eq:cx} is the Tafel rate of the ORR.
In a functionally graded CCL,
the parameters $\Cdl$, $\sion$ and $\Dox$ can be functions of
the coordinate $\tx$.

In \cite{Kulikovsky_24f}, the model based on the system of equations \eqref{eq:etax}, \eqref{eq:cx}
with the uniform $\sion$, $\Dox$ and $\Cdl$ has been developed. Here, we consider the general
case of the CCL with non-uniform parameters.

Linearization and Fourier-transform of Eqs.\eqref{eq:etax}, \eqref{eq:cx} yields
the system of equations for the small perturbation amplitudes $\teta^1$ and $\tc^1$:
\begin{multline}
    \pdr{}{\tx}\left(\tsion \pdr{\teta^1}{\tx}\right) = p_* \tc^1 + q_* \teta^1, \\
       \teta^1(1) = \teta_1^1, \quad \left.\pdr{\teta^1}{\tx}\right|_{\tx=1} = 0
    \label{eq:teta1x}
\end{multline}
\begin{multline}
    \pdr{}{\tx}\left(\tDox \pdr{\tc^1}{\tx}\right) = r_* \tc^1 + s_* \teta^1, \\
    \left.\pdr{\tc^1}{\tx}\right|_{\tx=0} = 0, \quad \tc^1(1) = \tc_1^1
    \label{eq:tc1x}
\end{multline}
Here, the superscripts 0 and 1 mark the static variables and the small perturbation
amplitudes, respectively, $\omega$ is the angular frequency of the applied AC signal, and
 the dimensionless variables are
\begin{multline}
    \tit = \dfrac{t}{t_*} , \quad \tx = \dfrac{x}{\lcat},
    \quad \tc = \dfrac{c}{\cref}, \quad \teta = \dfrac{\eta}{b_T},
    \quad \tj = \dfrac{j}{j_*},  \\
    \tCdl = \dfrac{\Cdl}{\Cdlast}, \quad \tsion = \dfrac{\sion}{\sionast},  \quad \tom = \omega t_*, \\
     \tDox = \dfrac{4 F \Dox \cref}{\sionast b_T}, \quad
     \tZ = \dfrac{Z\sionast}{\lcat}.
    \label{eq:dless}
\end{multline}
where $\sionast$ is the reference proton conductivity,
$\Cdlast$ is the reference DL capacitance,
and the characteristic time and current density are
\begin{equation}
   t_* = \dfrac{\Cdlast b_T}{i_*}, \quad j_* = \dfrac{\sionast b_T}{\lcat}.
   \label{eq:tjast}
\end{equation}

The coefficients in Eqs.\eqref{eq:teta1x} and \eqref{eq:tc1x} depend on the coordinate $\tx$
and frequency $\tom$:
\begin{align}
    p_* &= \dfrac{\eeo}{\veps^2},                &q_* &= \dfrac{\tc^0\eeo + \ri\tom\tCdl}{\veps^2},  \nonumber \\
    r_* &= \dfrac{\eeo + \ri\tom\mu^2}{\veps^2}, &s_* &= \dfrac{\tc^0\eeo}{\veps^2},
    \label{eq:pqrs}
\end{align}
where  $\veps$ and $\mu$ are the dimensionless parameters
\begin{equation}
    \veps = \sqrt{\dfrac{\sionast b_T}{i_*\lcat^2}}, \quad
    \mu = \sqrt{\dfrac{4 F \cref}{\Cdlast b_T}}.
    \label{eq:muveps}
\end{equation}
Note that the coefficients, Eq.\eqref{eq:pqrs}, differ
from those used in \cite{Kulikovsky_24f}.

The system  \eqref{eq:teta1x}, \eqref{eq:tc1x} contains the static
functions $\tc^0(\tx)$, $\teta^0(\tx)$, the
DL capacitance $\tCdl(\tx)$, proton conductivity $\tsion(\tx)$
and oxygen diffusivity $\tDox(\tx)$.
The calculation of the static shapes $\teta^0$ and $\tc^0$ is described
in Appendix~\ref{sec:anly}.

The system  \eqref{eq:teta1x}, \eqref{eq:tc1x} can be solved numerically by means
of a suitable BVP solver. However, this procedure is slow (see below). The matrix model
provides a much faster solution, as discussed in the next section.

\subsection{Matrix model for the functionally graded catalyst layer}

We introduce a numerical mesh on the interval $\tx \in [0, 1]$
with mesh nodes located at the centres of the numerical cells
(Figure~\ref{fig:sketch}). For the interval $[a_n, b_n]$, we consider
the Cauchy problem for the system of equations \eqref{eq:teta1x}, \eqref{eq:tc1x}
with ``frozen'' coefficients:
\begin{multline}
    \tsion \pddr{\teta^1}{\tx} = p_* \tc^1 + q_* \teta^1, \\
       \teta^1(b) = \teta_b^1, \quad \left.\tsion \pdr{\teta^1}{\tx}\right|_{\tx=b} = -\tj_b
    \label{eq:teta1x2}
\end{multline}
\begin{multline}
    \tDox \pddr{\tc^1}{\tx} = r_* \tc^1 + s_* \teta^1, \\
     \tc^1(b) = \tc_b^1, \quad \left.\tDox\pdr{\tc^1}{\tx}\right|_{\tx=b} = - N_b^1.
    \label{eq:tc1x2}
\end{multline}
Here, the parameters $\tsion$, $\tDox$, $\tCdl$, $\teta^0$, and $\tc^0$ are constants
calculated at $\tx_n$. However, they vary from
one numerical cell to another, i.e., their smooth dependencies
on $\tx$ are replaced by stepwise functions.

The system \eqref{eq:teta1x2}, \eqref{eq:tc1x2} with constant coefficients
can be solved analytically using any math software, e.g., \RMaple. Substituting $\tx=a$
into the solution yields linear relations between the solution vectors
on the right and left sides of the interval $[a, b]$ (here, the subscript $n$ is omitted)
\begin{equation}
\begin{pmatrix}
   &\teta_a^1 \\
   &\tc_a^1 \\
   &\tN_a^1  \\
   &\tj_a^1
\end{pmatrix}
=
\begin{pmatrix}
    &a_{11}      &a_{12}      &a_{13}      &a_{14}   \\
    &a_{21}      &a_{22}      &a_{23}      &a_{24}   \\
    &a_{31}      &a_{32}      &a_{33}      &a_{34}   \\
    &a_{41}      &a_{42}      &a_{43}      &a_{44}
\end{pmatrix}
\begin{pmatrix}
   &\teta_b^1 \\
   &\tc_b^1 \\
   &\tN_b^1  \\
   &\tj_b^1
\end{pmatrix}
   \label{eq:pmat}
\end{equation}
The matrix elements $a_{ij}$ are given in Appendix~\ref{sec:elem}.

Eqs.\eqref{eq:pmat} can be written in the matrix notations as
\begin{equation}
   \vecv_{n+1} = \matM_n \vecv_n
   \label{eq:rec}
\end{equation}
where $\matM_n$ is a $4\times 4$ matrix with components $a_{ij}$,
and $\vecv_n$, $\vecv_{n+1}$
are the right- and left-hand side vectors in Eqs.\eqref{eq:pmat}.

Starting with the vector $\vecv_1$ of parameters at $\tx=1$
\begin{equation}
   \vecv_1 = \left( \teta_1^1, \tc_1^1 , \tN_1^1, \tj_1^1 \right)^T
   \label{eq:v1}
\end{equation}
and applying Eq.\eqref{eq:rec} consecutively for $n=1, 2, \ldots, N$, we get
the vector $\vecv_{N+1}$ with the parameters at $\tx=0$:
\begin{equation}
   \vecv_{N + 1} = \matU \vecv_1
   \label{eq:vNp1}
\end{equation}
where the matrix $\matU$ is the product
\begin{equation}
   \matU = \matM_N\cdot \matM_{N-1} \ldots \matM_1
   \label{eq:matU}
\end{equation}
The components of the vector $\vecv_{N+1}$ are the values at the membrane
interface (Figure~\ref{fig:sketch}). The system impedance is given by
\begin{equation}
   \tZ =\dfrac{\teta_{N+1}^1}{\tj_{N+1}^1} = \dfrac{(\vecv_{N+1})_1}{(\vecv_{N+1})_4}
   \label{eq:tZ}
\end{equation}
Here, $(\vecv_{N+1})_k$ is the $k$th component of the vector $\vecv_{N+1}$.
Therefore, the calculation of impedance reduces to calculation of the matrix $\matU$ and
the product $\matU \vecv_1$.

\subsection{Initial conditions vector $\vecv_1$}

Two components of $\vecv_1$ in Eq.\eqref{eq:v1} are well defined:
$\tj_1^1 = 0$, and $\teta_1^1$ is an arbitrary small
applied perturbation amplitude. The components $\tc_1^1$ and $\tN_1^1$
need to be calculated.

To determine these components, we note that the oxygen concentration $\tc_1^1$
and flux $\tN_1^1$ on the cathode side of the CCL/GDL interface
must be equal to the concentration and flux on the GDL
side of this interface. An analytical solution of the transient
Ficks' equation for the oxygen transport  in the GDL leads to \cite{Kulikovsky_17i}
\begin{equation}
   \tc_1^1 - \alpha \tN_1^1 = \beta \tc_h^1
   \label{eq:Rob}
\end{equation}
where $\tc_h^1$ is the perturbation amplitude of the oxygen concentration in the channel, and
$\alpha$,  $\beta$ are the coefficients that take into account oxygen transport
in the GDL:
\begin{multline}
   \alpha = \dfrac{\tanh\left(\mu\tl_b\theta\right)}{\mu\tD_b\theta}, \quad
                  \beta = \dfrac{1}{\cosh\left(\mu\tl_b\theta\right)}, \\
                  \theta \equiv \sqrt{\ri\tom/\left(\veps^2\tD_b\right)}
   \label{eq:albeta}
\end{multline}
Here,
$\tD_b$ is the GDL oxygen diffusivity and $\tl_b$ thickness.

The relation between $\tc_h^1$ and $\tN_1^1$ results from Eq.\eqref{eq:vNp1}.
The third component of the vector $\vecv_{N+1}$ is the oxygen flux at the membrane surface.
This component must be zero, and we write
\begin{equation}
     (\vecv_{N+1})_3 = U_{31} \teta_1^1 + U_{32}\tc_1^1 + U_{33} \tN_1^1  = 0
     \label{eq:U1}
\end{equation}

Solving Eqs.\eqref{eq:Rob} and \eqref{eq:U1} for $\tN_1^1$, we get the oxygen
flux perturbation at the CCL/GDL interface
\begin{equation}
   \tN_1^1 = - \dfrac{U_{32}\beta\tc_h^1 + U_{31}\teta_1^1}{\alpha U_{32} + U_{33}}.
   \label{eq:tN11sol}
\end{equation}
Eqs.\eqref{eq:Rob} and \eqref{eq:tN11sol} include the oxygen concentration perturbation
in the channel $\tc_h^1$, which has to be determined.

\subsection{Oxygen mass transport in the channel}

To a good approximation, oxygen mass transport in the air channel can be
described by the 1d + 1d plug flow equation:
\begin{equation}
   \pdr{c_h}{t} + v\pdr{c_h}{z} = - \left.\dfrac{D_b}{h}\pdr{c_b}{x}\right|_{x=\lcat + l_b}
   \label{eq:chz}
\end{equation}
where
$c_h$ is the oxygen concentration in channel,
$v$ the flow velocity, and
$h$ the channel depth.
The right side of Eq.\eqref{eq:chz} is the oxygen diffusive flux
in the GDL at the channel/GDL interface, which represents the
oxygen ``sink''.

The Fourier-transform of Eq.\eqref{eq:chz} and the solution of the oxygen transport
problem in the GDL lead to the following equation for the
small perturbation amplitude $\tc_h^1$ (Ref.\cite{Kulikovsky_17i})
\begin{equation}
   \lam\tJ \pdr{\tc_h^1}{\tz} =  -\zeta\tc_h^1
             + \beta\tN_1^1, \quad \tc_h^1(0) = 0
   \label{eq:tch1z}
\end{equation}
where we introduced a notation
\begin{equation}
   \zeta \equiv  \ri\tom\xi^2 + \mu \sqrt{\ri\tom\tD_b/\veps^2}\,\tanh\left(\mu\tl_b\sqrt{\sqa}\right).
   \label{eq:zeta}
\end{equation}
Here, $\beta$ is given by Eq.\eqref{eq:albeta}, $\xi$ is the constant parameter, and
$\lam$ is the stoichiometry of the air flow
\begin{equation}
    \xi = \sqrt{\dfrac{4 F h \cref\lcat i_*}{\Cdlast\sionast b_T^2}},\quad \lam = \dfrac{4 F h v \cref}{L J}.
    \label{eq:xilam}
\end{equation}

The solution to Eq.\eqref{eq:tch1z} contains complex exponent, which
rapidly oscillates along $\tz$ (see below).
This feature makes the procedure of
accurate numerical solution to Eq.\eqref{eq:tch1z} extremely time-consuming. The
initial-value solver reduces the step size along $\tz$ to very small values in order to
resolve the rapidly oscillating solution.

A good analytical approximation
to Eq.\eqref{eq:tch1z} solution is obtained as follows.
Eq.\eqref{eq:tch1z} contains the through-plane oxygen flux at the CCL/GDL interface
$\tN_1^1 = - \tDox\pdra{\tc^1}{\tx}|_{\tx=1}$, which depends on the coordinate $\tz$.
However, this dependence is weak and to a zero-order approximation
we solve Eq.\eqref{eq:tch1z}, assuming that $\tN_1^1$ is constant:
\begin{equation}
   \tc_h^1(\tz) = \dfrac{\beta \tN_1^1}{\zeta}
                  \left(1 - \lexp{- \dfrac{\zeta\tz}{\lam\tJ}} \right)
   \label{eq:tch1_sol}
\end{equation}
Substituting Eq.\eqref{eq:tch1_sol} into Eq.\eqref{eq:tN11sol}
and solving the resulting equation for $\tN_1^1$,
we find a more accurate, explicit dependence of $\tN_1^1$ on
the distance along the channel $\tz$:
\begin{equation}
\small
   \tN_1^1 = - \dfrac{U_{31}\teta_1^1 }
                     {\left(\dfrac{\beta^2}{\zeta}
                      \left(1 - \lexp{- \dfrac{\zeta\tz}{\lam\tJ}}\right)
                       + \alpha\right) U_{32} + U_{33}}.
    \label{eq:tN11_sol}
\end{equation}
Finally, we substitute Eq.\eqref{eq:tN11_sol} into Eq.\eqref{eq:tch1_sol},
and determine $\tc_1^1$ using Eq.\eqref{eq:Rob}. These analytical
iterations yield the missing
components $\tc_1^1$ and $\tN_1^1$ of the vector $\vecv_1$.



\subsection{Numerical implementation}

The procedure of spectra calculation using the matrix model includes the following steps.

\begin{enumerate}

\item Separate the channel coordinate $\tz \in [0, 1]$
     into $N_{seg}$ segments. For each segment, perform steps \#2 to  \#6 below.

\item Using any BVP solver, solve the system of static equations \eqref{eq:teta0x},
  \eqref{eq:tc0x}.

\item Introduce a numerical mesh on the interval $\tx \in [0,1]$  with $N$ nodes located
      in the middle of $[a_n, b_n]$: $\tx_n = (a_n + b_n) / 2, n=1, \ldots, N$
      (Figure~\ref{fig:sketch}). Note that $n=1$ corresponds to the rightmost
      cell (at $\tx=1$).

\item In a loop over $n=1, \ldots, N$ perform the following operations for all the frequencies
    \begin{itemize}
        \item Calculate the coefficient functions $p_*, q_*, r_*, s_*$, Eq.\eqref{eq:pqrs}.

        \item Using the equations in the Appendix~\ref{sec:elem}, calculate the coefficients of the matrix $\matM_n$.

        \item Calculate the matrix $\matU$ by accumulating the product of the
              matrices $\matM_n$, Eq.\eqref{eq:matU}.

    \end{itemize}

\item Using Eqs.\eqref{eq:tN11_sol}, \eqref{eq:tch1_sol},  and \eqref{eq:Rob}, calculate the
      parameters $\tN_1^1$ and $\tc_1^1$ of the vector $\vecv_1$.

\item Calculate the vector $\vecv_{N+1}$ from Eq.\eqref{eq:vNp1} and
      the segment impedance $\tZ_{seg}$ from Eq.\eqref{eq:tZ}.

\item Once all segment impedances have been obtained, calculate the cell impedance $\tZ$ from the equation
   \begin{equation}
      \dfrac{1}{\tZ} = \dfrac{1}{N_{seg}}\sum_{k=1}^{N_{seg}}\dfrac{1}{\tZ_{seg, k}}
      \label{eq:tZsys}
   \end{equation}

\end{enumerate}
The code can easily be parallelized into $N_{seg}$ processes, with each process
calculating the impedance $\tZ_{seg}$ of
individual segments. Typically, $N_{seg} = 8$ is sufficient to
describe the non-uniform distribution of the local current and
oxygen concentration along the cathode channel.

The slowest operation is calculating the matrix $\matU$, which involves multiplying
$N$ $4\times 4$ matrices for each frequency. In Python, this procedure can be performed
using a call to {\em np.matmul}, which enables batch multiplication of matrices.
The Python program for the matrix model can thus  be constructed without
a single loop over the frequencies, resulting in highly efficient code.
In contrast, the Python code based on the BVP solver {\em solve\_bvp} is slow due
to (i) internal iterations in the BVP solver,
(ii) a loop over the frequencies,
and (iii) iterations to derive a self-consistent solution to Eq.\eqref{eq:tch1z}.
{\em The calculations show that the matrix model outperforms the BVP-solver model
by nearly three orders of magnitude.}

\section{Results and discussion}

In this section, the reference spectra were obtained using
the Python BVP solver for the system of equations \eqref{eq:teta1x}, \eqref{eq:tc1x}.
Below, this model will
be referred to as the reference one. The set of parameters used
in the calculations is listed in Table~\ref{tab:parms}.
Note that the BVP-solver model does not include the matrix $\matU$ and employs
a numerical iterative solution to Eq.\eqref{eq:tch1z}. Below, the matrix model
is compared with the BVP-solver model, which uses an approximate numerical
solution to Eq.\eqref{eq:tch1z}. This solution is obtained assuming
that the flux $\tN_1^1$ is constant along each segment.

\begin{figure}
\begin{center}
    \includegraphics[scale=0.45]{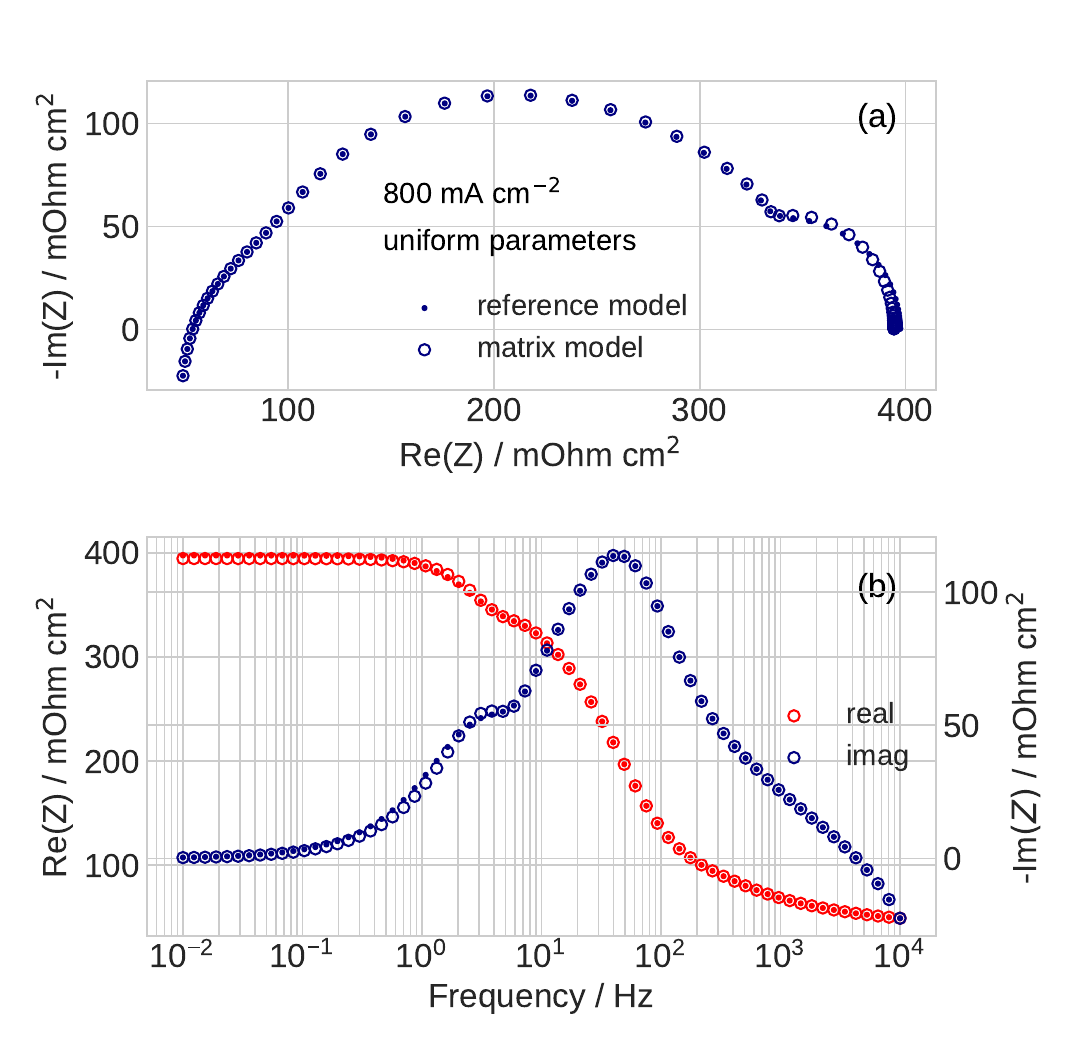}
\end{center}
\caption{(a) The Nyquist and (b) Bode plots of the PEMFC cathode impedance
    calculated using the reference model (points) and the matrix model (open circles)
    with the numerical mesh containing $N=20$ points.
    The CCL transport parameters are uniform.
    The matrix model returns the spectrum 600 times faster, than the reference model.
    }
\label{fig:s800u}
\end{figure}

Figure~\ref{fig:s800u} compares the impedance spectra of a PEMFC
calculated using the reference and matrix models. The transport
parameters are uniform through the CCL depth and the cell current density
is 800 mA~cm$^{-2}$.  Note the low-frequency arc due to
the oxygen transport in the channel \cite{Ingo_07a}.
As can be seen, the matrix model with $N=20$ numerical cells
demonstrates an excellent accuracy. Note that the reference model
could actually be less accurate than the matrix one.
The reason is that the reference model employs an approximate
numerical solution to Eq.\eqref{eq:tch1z}.

\begin{table}
\small
\begin{center}
\begin{tabular}{|l|c|}
        \hline
        CCL thickness $\lcat$, cm            &  $10\cdot 10^{-4}$  \\
        CCL oxygen diffusivity $\Dox$, cm$^2$s$^{-1}$ &  $10^{-4}$  \\
        ORR Tafel slope, V / exp             &  0.03     \\
        ORR exchange current density         &          \\
        $i_*$, A~cm$^{-3}$                   &  $10^{-4}$   \\
        Double layer capacitance $\Cdlast$, F~cm$^{-3}$  & 20  \\
        CCL proton conductivity $\sionast$, S~cm$^{-1}$  & $5\cdot 10^{-3}$ \\
        GDL thickness $l_b$, cm              &   $235\cdot 10^{-4}$  \\
        GDL oxygen diffusivity $D_b$, cm$^2$s$^{-1}$ &  $0.02$  \\
        Number of segments $N_{seg}$         &   8   \\
        \hline
        Cathode absolute pressure, bar       &  1     \\
        Cathode flow RH                      &  0.5   \\
        Air flow stoichiometry $\lam$        &  3     \\
        Cell temperature, K                  &  273 + 80 \\
        Cell current density $j_0$, mA~cm$^{-2}$   & 800 \\
        \hline
\end{tabular}
\end{center}
\caption{Cell geometrical, transport and operating parameters.
}
    \label{tab:parms}
\end{table}

\begin{figure}
\begin{center}
    \includegraphics[scale=0.45]{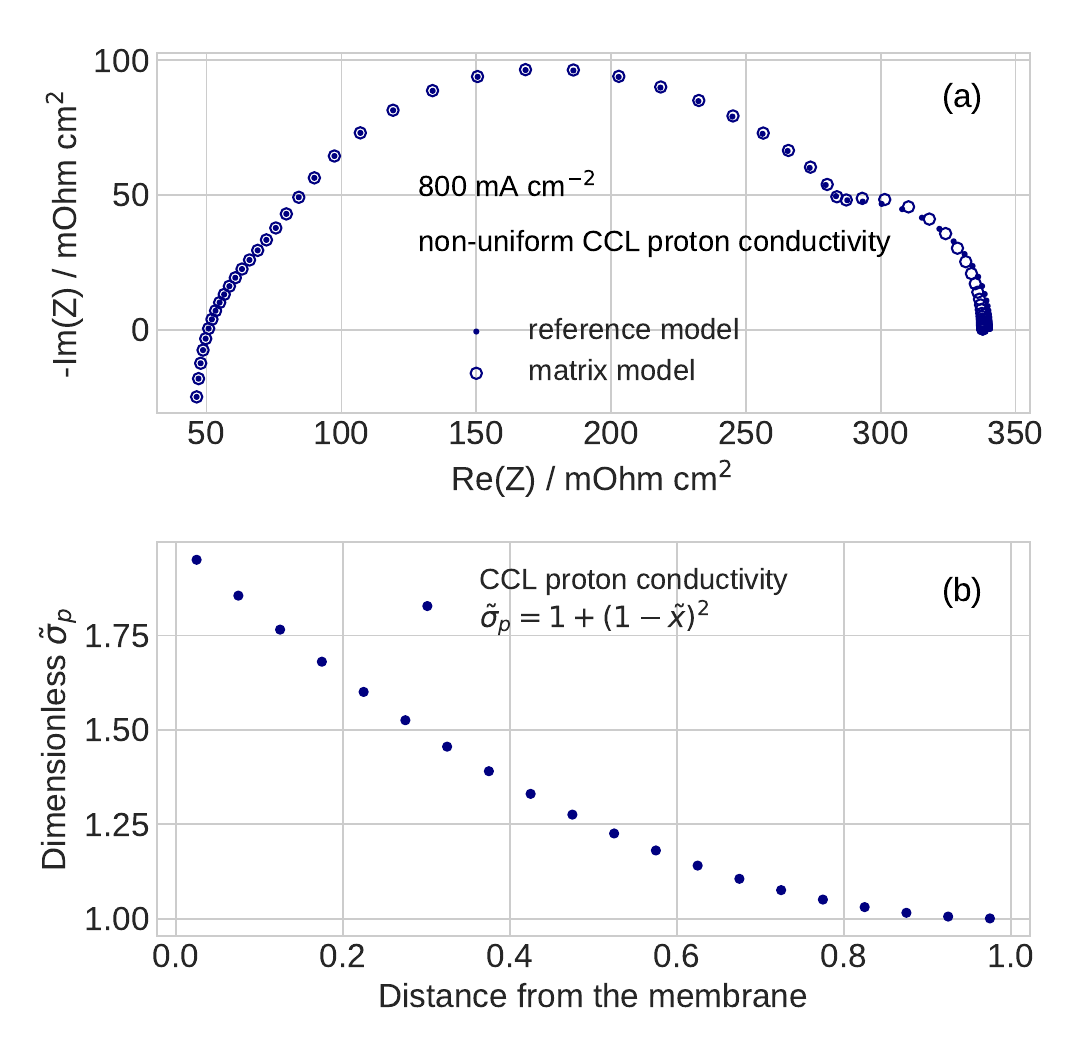}
\end{center}
\caption{(a) The Nyquist plot of the PEMFC cathode impedance
    calculated using the reference (points) and the matrix model (open circles)
    with $N=20$ points. (b) The shape of the CCL proton conductivity used in the calculations.
    The matrix model returns the spectrum 600 times faster, than the reference model.
    }
\label{fig:s800sig}
\end{figure}

\begin{figure}
\begin{center}
    \includegraphics[scale=0.45]{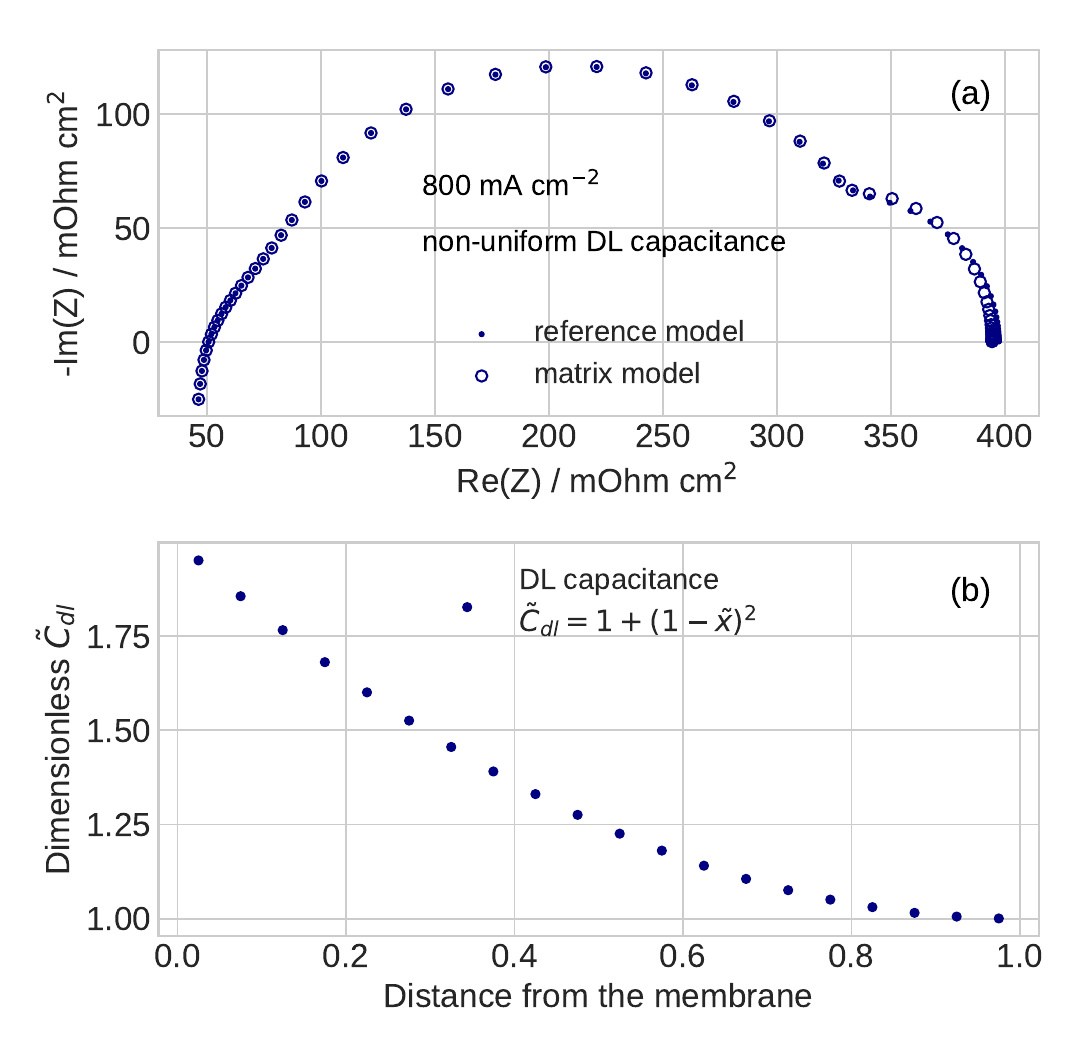}
\end{center}
\caption{(a) The Nyquist plot of the PEMFC cathode impedance
    calculated using the reference (points) and the matrix model (open circles)
    with $N=20$ points. (b) The shape of the double layer capacitance used in the calculations.
    The matrix model returns the spectrum 600 times faster, than the reference model.
    }
\label{fig:s800cdl}
\end{figure}

Figure~\ref{fig:s800sig}a shows the spectra calculated using the reference
and matrix models for the case of a non-uniform CCL proton conductivity
(Figure~\ref{fig:s800sig}b). Figure~\ref{fig:s800cdl} shows the spectra
for the non-uniform DL capacitance. Comparing Figure~\ref{fig:s800cdl}a with
Figure~\ref{fig:s800u}a, a higher slope of the Nyqist spectrum in Figure~\ref{fig:s800cdl}a in the
high-frequency range can be seen. This effect is caused by a non-uniform $\Cdl$.
Both Figures~\ref{fig:s800sig} and \ref{fig:s800cdl} show a high accuracy of the matrix model.


\begin{figure}
\begin{center}
    \includegraphics[scale=0.45]{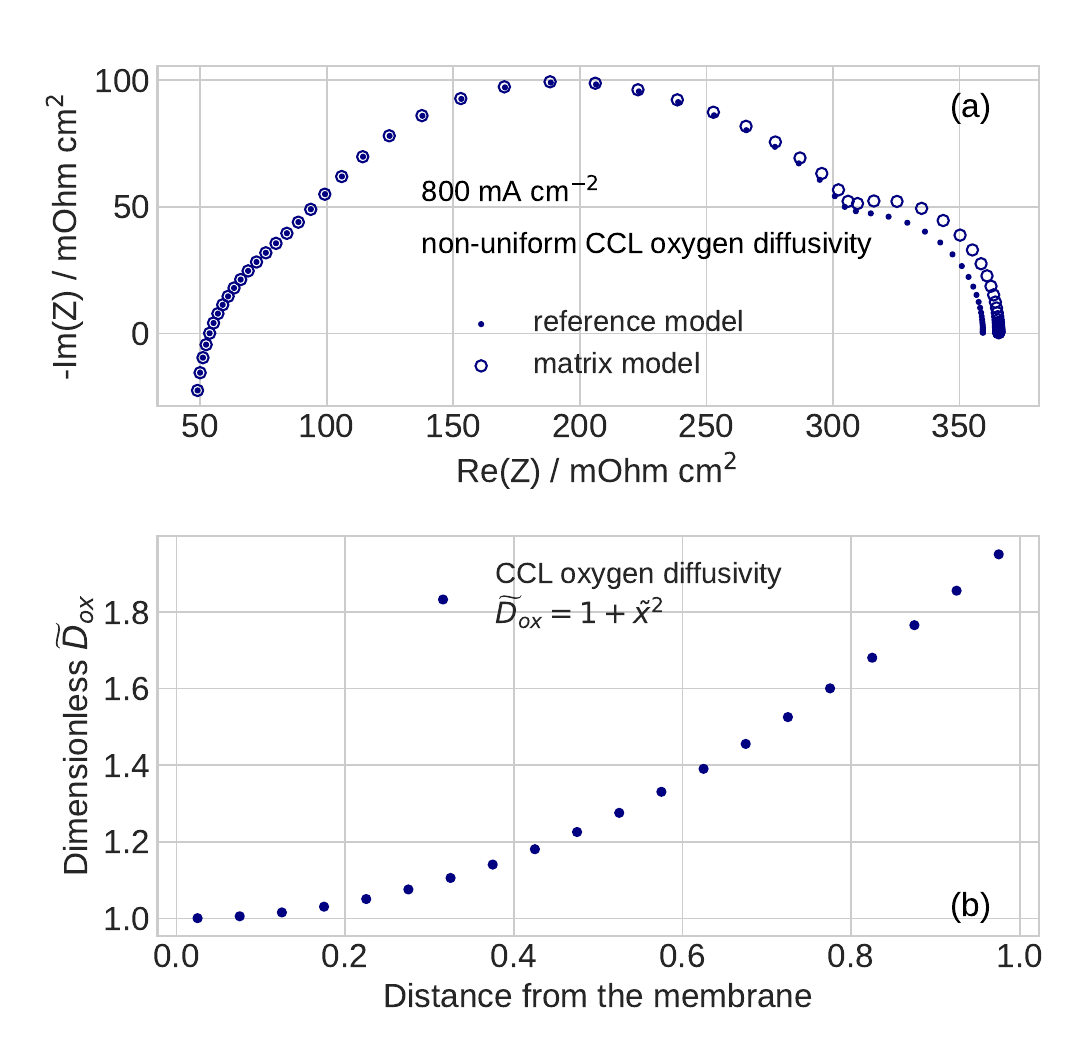}
\end{center}
\caption{(a) The Nyquist plot of the PEMFC cathode impedance
    calculated using the reference (points) and the matrix model (open circles)
    with $N=20$ points. (b) The shape of the CCL oxygen diffusivity used in the calculations.
    }
\label{fig:s800dox}
\end{figure}

Figure~\ref{fig:s800dox}a compares the spectra for the case of a non-uniform CCL oxygen diffusivity
(Figure~\ref{fig:s800dox}b).
There is a small gap between the two spectra in
the low-frequency region (Figure~\ref{fig:s800dox}a).
This gap is presumably due to insufficient accuracy of the iterative solver for Eq.\eqref{eq:tch1z}
in the reference model. At small currents, the gap between the Nyquist curves vanishes.

\begin{figure}
\begin{center}
    \includegraphics[scale=0.45]{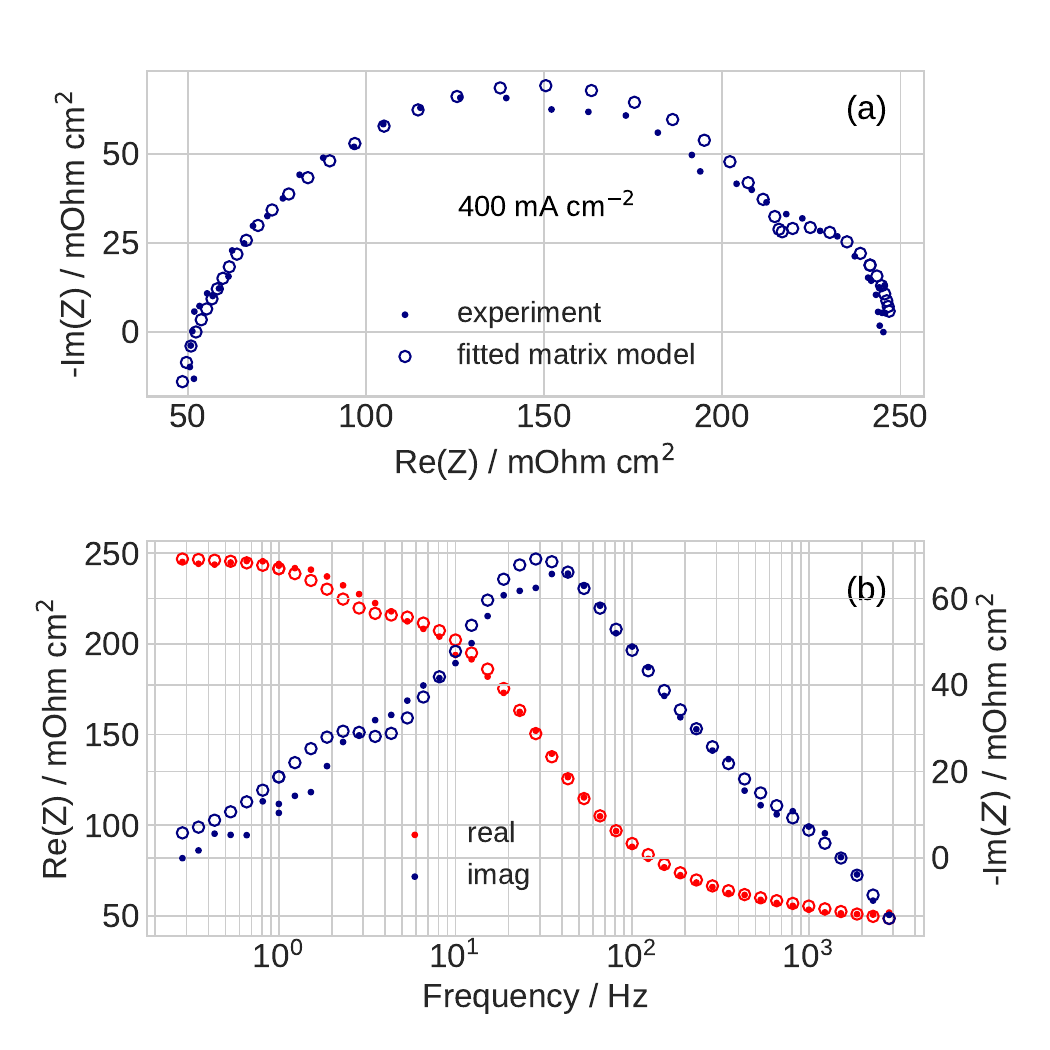}
\end{center}
\caption{(a) The experimental (points) and fitted matrix model (open circles)
    Nyquist plots of the PEMFC impedance. The number of mesh points is $N=20$;
    the CCL transport parameters are assumed to be uniform. (b) The Bode plots of the real
    and imaginary part of the impedances in (a). The fitted parameters
    are listed in Table~\ref{tab:fitted}. The fitting takes 2 s on a standard notebook.
    }
\label{fig:S067_800}
\end{figure}

Figure~\ref{fig:S067_800} shows the fitting of the experimental PEMFC spectrum using the matrix model.
The cell operating parameters are listed in Table~\ref{tab:oper}; more experimental details can be found in\cite{Kulikovsky_16d}.

\begin{table}
\small
\begin{center}
\begin{tabular}{|l|c|}
        \hline
        CCL thickness $\lcat$, cm            &  { $10\cdot 10^{-4}$}   \\
        GDL thickness $l_b$, cm              &  { $235\cdot 10^{-4}$}  \\
        Air flow stoichiometry $\lambda$     &  4   \\
        Cathode absolute pressure, bar       &  1.5   \\
        Cathode flow RH                      & 0.5   \\
        Cell temperature, K                  &  $273 + 80$ \\
        Cell current density $j_0$, mA~cm$^{-2}$   & 400 \\
        ORR exchange current density,  $i_*$, A~cm$^{-3}$ &  $10^{-4}$   \\
        Cell active area $S$, cm$^2$         &  76 \\
        \hline
\end{tabular}
\end{center}
\caption{Cell operating parameters.
}
   \label{tab:oper}
\end{table}
%




%
\begin{table}
\small
\begin{center}
\begin{tabular}{|l|c|}
        \hline
        ORR Tafel slope, mV / exp                      & $31 \pm 3$   \\
        Double layer capacitance $\Cdlast$, F~cm$^{-3}$   & $20 \pm 2$   \\
        CCL proton conductivity $\sion$, mS~cm$^{-1}$  & $16 \pm 3$   \\
        CCL oxygen diffusivity $\Dox$, cm$^2$s$^{-1}$  & $(0.9 \pm 0.2)\cdot 10^{-4}$  \\
        GDL oxygen diffusivity $D_b$, cm$^2$s$^{-1}$   & $(3.1 \pm 0.5)\cdot 10^{-2}$ \\
        HFR, mOhm~cm$^2$                               & $39 \pm 2$   \\
        Cable inductance, $L_{cab}$, nH                & $17 \pm 2$   \\
        \hline
        Fitting code runtime, s                        & 2.0   \\
        \hline
\end{tabular}
\end{center}
\caption{Fitted parameters for the spectrum in Figure~\ref{fig:S067_800} and the runtime
         of the fitting code.
}
   \label{tab:fitted}
\end{table}

The quality of fitting the low-frequency range of the spectrum is not perfect.
However, the fitted parameters (Table~\ref{tab:fitted}) are within the typical range for
a standard Pt/C-based PEMFC \cite{Kulikovsky_26b}. Note that fitting of this spectrum takes
about 2 s on a standard notebook.

Finally, we note that the only problem requiring a classic collocation-type boundary
value solver in the matrix model
is the static equations, Eq.\eqref{eq:teta0x}, \eqref{eq:tc0x}. This procedure is fast, as the
static shapes $\teta^0(\tx)$ and $\tc^0(\tx)$ typically are smooth functions.

\section{Conclusions}

We have developed a fast, physics-based matrix model for calculating the impedance
of PEM fuel cells. The model is suitable for PEMFCs with a functionally graded
cathode catalyst layer. It also takes into account the finite stoichiometry
of the air (oxygen) flow in the channel.
Our model is almost three orders of magnitude faster than the model based on
the standard collocation method for numerically solving the linear BVP
for perturbation amplitudes.  It can be used for fast spectra fitting in on-board
testing systems and for EIS-based analysis of cells with functionally
graded catalyst layers.

\appendix

\section{Static equations}
\label{sec:anly}

The parameters $p_*, q_*, r_*, s_*$, Eq.\eqref{eq:pqrs}, contain the static overpotential $\teta^0$
and oxygen concentration $\tc^0$, which obey the system of equations
\begin{equation}
   \veps^2 \pddr{\teta^0}{\tx} = \tc^0 \expn,
      \quad \left.\pdr{\teta^0}{\tx}\right|_{\tx=0} = - \tj_0,
      \quad \left.\pdr{\teta^0}{\tx}\right|_{\tx=1} = 0
   \label{eq:teta0x}
\end{equation}
\begin{equation}
     \veps^2\tDox \pddr{\tc^0}{\tx} = \tc^0\expn,  \quad
     \left.\pdr{\tc^0}{\tx}\right|_{\tx=0} = 0, \quad \tc^0(1) = \tc_1^0
     \label{eq:tc0x}
\end{equation}
Generally, the solution of this strongly nonlinear problem can only be obtained numerically,
using any standard BVP solver.
However, if the oxygen transport losses in the CCL are small, an analytical
solution to Eq.\eqref{eq:teta0x} can be used.

If the oxygen concentration gradient through the CCL is small,
we cat set $\tc^0 \simeq \tc_1^0$ In Eq.\eqref{eq:teta0x},
where $\tc_1^0$ is the concentration at $\tx=1$.  In this case,
the static overpotential $\teta^0$ obeys the equation
\begin{equation}
   \veps^2 \pddr{\teta^0}{\tx} = \tc_1^0 \exp\teta^0,
      \quad \left.\pdr{\teta^0}{\tx}\right|_{\tx=0} = -\tj_0,
      \quad \left.\pdr{\teta^0}{\tx}\right|_{\tx=1} = 0
   \label{eq:teta0x1}
\end{equation}
The solution to Eq.\eqref{eq:teta0x1} is \cite{Neyerlin_07a,Kulikovsky_10c}
\begin{equation}
    \teta^0(\tx) = \lnl{\dfrac{\veps^2\beta^2}{2\tc_1^0}
                \left(1 + \tan^2\left(\dfrac{\beta}{2}\left(1 - \tx\right)\right)\right)}
    \label{eq:teta0x_sol}
\end{equation}
where
$\beta$ is a solution to the equation
\begin{equation}
    \beta\tan\left(\dfrac{\beta}{2}\right) = \tj_0, \quad 0 \leq \beta < \pi
    \label{eq:beta}
\end{equation}
and $\tj_0$ is the static local current density.

\begin{figure}
 \begin{center}
   \includegraphics[scale=0.45]{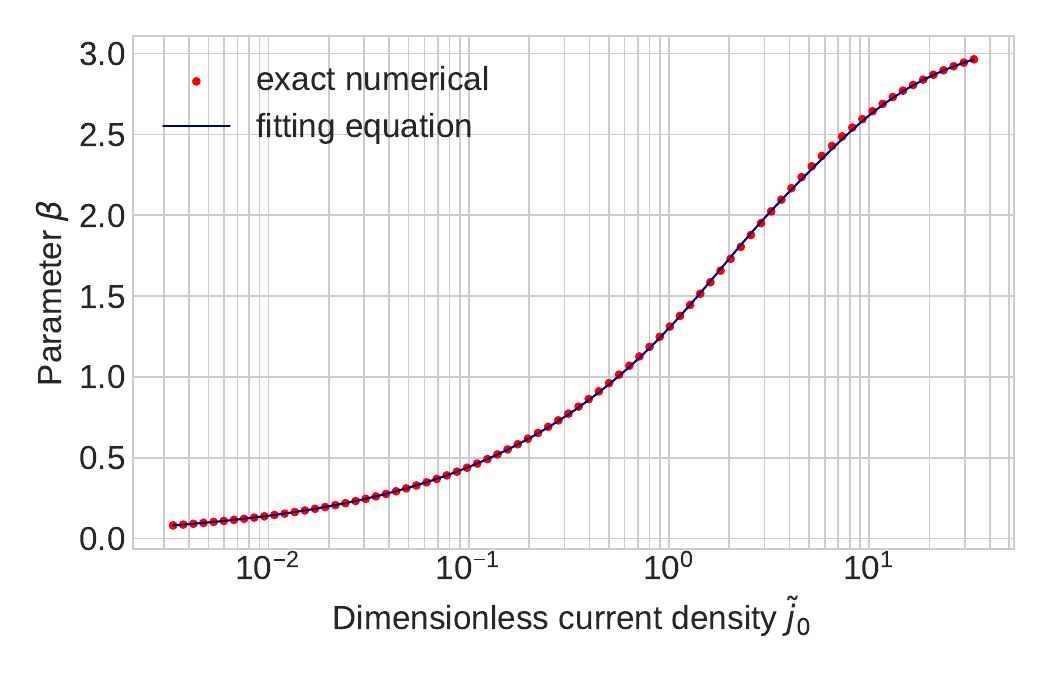}
\end{center}
   \caption{The numerical solution to Eq.\eqref{eq:beta} (points) and the best-fit
   matching function Eq.\eqref{eq:beta_apx}
   with  $a = 1.428$,  $k = 0.3645$ (solid line).
   }
    \label{fig:beta}
\end{figure}

For small and large $\tj_0$, the asymptotic solutions to Eq.\eqref{eq:beta} are
\cite{Kulikovsky_12g}
\begin{equation}
   \beta = \left\{ \begin{split}
                    &\sqrt{2\tj_0}, \quad \tj_0 \ll 1, \\
                    & \dfrac{\pi \tj_0}{2 + \tj_0}, \quad \tj_0 \gg 1
                   \end{split}
    \right.
    \label{eq:limits}
\end{equation}
The best-fit matching function, which links the two asymptotics is
\begin{equation}
   \beta = \lexp{-k \tj_0^a}\sqrt{2\tj_0} + \left(1  - \lexp{-k \tj_0^a}\right) \dfrac{\pi \tj_0}{2 + \tj_0}
   \label{eq:beta_apx}
\end{equation}
where $a \simeq 1.428$,  $k \simeq 0.3645$ (Figure~\ref{fig:beta}).

%

The local current density $\tj_0$ and oxygen
concentration $\tc_1^0$ are the functions of the coordinate $\tz$ along the channel.
The shape of $\tj_0(\tz)$ can be obtained using the model \cite{Kulikovsky_19a}.
The parameter $\tc_1^0$ is the solution to equations
\begin{equation}
   \lam\tJ \pdr{\tc_h^0}{\tz} = -\tj_0(\tz), \quad
       \tc_1^0(\tz) = \tc_h^0(\tz) - \dfrac{\tj_0(\tz)}{\tjlim}
   \label{eq:tch0z}
\end{equation}
where $\tJ$ is the mean current density in the cell, $\lam$ is the air flow stoichiometry,
the dimensionless distance $\tz = z/L$, $L$ is the channel length, and
\begin{equation}
   \jlim = \dfrac{4 F c_h^0 D_b}{l_b}
   \label{eq:jlim}
\end{equation}
is the limiting current density.

\section{Elements of the matrix $\matM_n$ in Eq.\eqref{eq:pmat}}
\label{sec:elem}

In this Section, $p, q, r, s$ are related to $p_*, q_*, r_*, s_*$,
Eqs.\eqref{eq:pqrs}, as
\begin{equation}
   p = \dfrac{p_*}{\tsion}, \quad q = \dfrac{q_*}{\tsion},
   \quad r = \dfrac{r_*}{\tDox}, \quad s = \dfrac{s_*}{\tDox}.
   \label{eq:pqrs_new}
\end{equation}
Further, $d_n = b_n - a_n$ is the length of the interval $[a_n, b_n]$ (Figure~\ref{fig:sketch}),
and the following notations are used:
\begin{equation}
    \begin{split}
        & \psi   = \sqrt{4p s + (q - r)^2}   \\
        & \phi_1 = \sqrt{2(q + r) + 2\psi}, \quad \phi_2 = \sqrt{2(q + r) - 2\psi}  \\
        & \xi_1  = \psi + q - r , \quad \xi_2  = \psi - q + r, \\
        & C_1 = \lcosh{\dfrac{\phi_1 d_n}{2}}, \quad C_2 = \lcosh{\dfrac{\phi_2 d_n}{2}} \\
        & S_1 = \lsinh{\dfrac{\phi_1 d_n}{2}}, \quad S_2 = \lsinh{\dfrac{\phi_2 d_n}{2}} \\
    \end{split}
    \label{eq:phi12}
\end{equation}
The parameters $p, q, r, s$ should be calculated for each interval $[a_n, b_n]$, as well
as the parameters in Eq.\eqref{eq:phi12} and the elements of the
 matrix in Eq.\eqref{eq:pmat}:
\begin{equation}
\begin{split}
    a_{11} &= \dfrac{1}{2\psi}\left(\xi_1C_1 +  \xi_2 C_2 \right), \quad
    a_{12} =  \dfrac{p}{\psi}\left(C_1 - C_2\right) \\
    a_{13} &= \dfrac{2p}{\phi_1\phi_2\psi\tDox} \left(\phi_2 S_1 - \phi_1 S_2 \right) \\
    a_{14} &= \dfrac{1}{\phi_1\phi_2\psi\tsion} \left(\phi_2\xi_1 S_1 + \phi_1\xi_2 S_2\right)
\end{split}
    \label{eq:a1}
\end{equation}

\begin{equation}
\begin{split}
    a_{21} &= \dfrac{s}{\psi}\left(C_1 - C_2\right), \quad
    a_{22}  = \dfrac{1}{2\psi}\left(\xi_1 C_2 + \xi_2C_1 \right) \\
    a_{23} &= \dfrac{1}{\phi_1\phi_2\psi\tDox}
             \left(\phi_1\xi_1S_2 + \phi_2\xi_2S_1 \right)  \\
    a_{24} &= - \dfrac{2 s}{\phi_1\phi_2\psi\tsion}\left(\phi_1 S_2 - \phi_2 S_1 \right)
\end{split}
   \label{eq:a2}
\end{equation}

\begin{equation}
\begin{split}
   a_{31} &= \dfrac{s\tDox}{2\psi} \left(\phi_1 S_1 - \phi_2 S_2 \right)   \\
   a_{32} &= \dfrac{\tDox}{4\psi}\left(\phi_2 \xi_1 S_2 + \phi_1 \xi_2 S_1 \right) \\
   a_{33} &= a_{22}, \quad
   a_{34}  = \dfrac{s\tDox}{\psi\tsion} \left(C_1 - C_2\right) \\
\end{split}
\label{eq:a3}
\end{equation}

\begin{equation}
\begin{split}
   a_{41} &=  \dfrac{\tsion}{4\psi} \left(\phi_1\xi_1 S_1 + \phi_2\xi_2 S_2\right) \\
   a_{42} &=  \dfrac{\tsion p}{2\psi}\left( \phi_1 S_1 - \phi_2 S_2\right)  \\
   a_{43} &=  \dfrac{\tsion p}{\psi\tDox}\left(C_1 - C_2\right),  \quad
   a_{44}  =  a_{11}
\end{split}
    \label{eq:a4}
\end{equation}

\newpage


\newpage

\section*{Nomenclature}

\small

\begin{tabular}{ll}
    $\tilde{}$   &  Marks dimensionless variables                          \\
    $b$          &  ORR Tafel slope, V                                     \\
    $d$          &  Dimensionless length of the interval $[a, b]$           \\
    $c$          &  Oxygen concentration in the CCL, mol~cm$^{-3}$          \\
    $c_b$        &  Oxygen concentration in the GDL, mol~cm$^{-3}$          \\
    $c_h$        &  Oxygen concentration in the channel, mol~cm$^{-3}$      \\
    $C_1, C_2$   &  Dimensionless functions, Eq.\eqref{eq:phi12}            \\
    $\cref$      &  Reference oxygen concentration, mol~cm$^{-3}$    \\
    $\Cdl$       &  Double layer volumetric capacitance, $F$~cm$^{-3}$      \\
    $D_b$        &  GDL oxygen diffusivity, cm$^2$~s$^{-1}$                \\
    $\Dox$       &  CCL oxygen diffusivity, cm$^2$~s$^{-1}$                \\
    $F$          &  Faraday constant, C~mol$^{-1}$                         \\
    $h$          &  Channel depth, cm                                      \\
    $\ri$        &  Imaginary unit                                         \\
    $i_*$        &  ORR volumetric exchange current density, A~cm$^{-3}$   \\
    $j_0, j^0$   &  Static current density, A~cm$^{-2}$                    \\
    $\lcat$      &  CCL thickness, cm                                      \\
    $l$          &  Sub-layer length, $l = \lcat/N$                        \\
    $l_b$        &  GDL thickness, cm                                      \\
    $\matM$      &  Matrix with the components Eq.\eqref{eq:pmat}           \\
    $N$          &  Number of computational cells in the CCL               \\
    $N_b, N_n$   &  Oxygen flux at the half-cell, mol~cm$^{-2}$~s$^{-1}$   \\
    $N_{seg}$    &  Number of segments                                     \\
    $\matU$      &  Matrix, Eq.\eqref{eq:matU}                             \\
    $p, q, r, s$ &  Coefficient functions, Eq.\eqref{eq:pqrs_new}          \\
    $p_*, q_*, r_*, s_*$ &  Coefficient functions, Eq.\eqref{eq:pqrs}      \\
    $S_1, S_2$   &  Dimensionless functions, Eq.\eqref{eq:phi12}            \\
    $\vecv_n$        &  Four-component dimensionless vector                    \\
    $x$          &  Coordinate through the cell, cm                        \\
    $z$          &  Coordinate along the channel, cm                       \\
    $Z_*$        &  Scaling factor for the impedance,  Ohm~cm$^2$          \\
    $Z$          &  Impedance, Ohm~cm$^2$               \\
    $Z_{seg}$    &  Segment impedance, Ohm~cm$^2$       \\[1em]
\end{tabular}

{\bf Subscripts:\\}

\begin{tabular}{ll}
    $a$      & Left side of the interval $[a, b]$ \\
    $b$      & Right side of the interval $[a, b]$ \\
    $0$      & membrane/CCL interface \\
    $1$      & CCL/GDL interface  \\
    $n$      & $n$th computational cell   \\
    $seg$    & Segment \\ [1em]
\end{tabular}

{\bf Superscripts:\\}

\begin{tabular}{ll}
    $0$      & Steady--state value \\
    $1$      & Small--amplitude perturbation \\ [1em]
\end{tabular}

\newpage

{\bf Greek:\\}

\begin{tabular}{ll}
    $\alpha$            &  Dimensionless parameter, Eq.\eqref{eq:albeta}  \\
    $\beta$             &  Dimensionless parameter, Eq.\eqref{eq:albeta}  \\
    $\veps$             &  Newman's dimensionless reaction  \\
                        &  penetration depth, Eq.\eqref{eq:muveps} \\
    $\zeta$             &  Dimensionless parameter, Eq.\eqref{eq:zeta}  \\
    $\eta$              &  ORR overpotential, positive by convention, V \\
    $\theta$            &  Dimensionless parameter, Eq.\eqref{eq:albeta}  \\
    $\lam$              &  Air flow (oxygen) stoichiometry, Eq.\eqref{eq:xilam}   \\
    $\mu$               &  Dimensionless parameter, Eq.\eqref{eq:muveps}  \\
    $\xi$               &  Dimensionless parameter, Eq.\eqref{eq:phi12}  \\
    $\xi_1, \xi_2$      &  Dimensionless parameters, Eq.\eqref{eq:xilam}  \\
    $\sion$             &  CCL proton conductivity, $\Omega^{-1}$~cm$^{-1}$        \\
    $\phi$              &  Dimensionless parameter, Eq.\eqref{eq:phi12}  \\
    $\psi$              &  Dimensionless parameter, Eq.\eqref{eq:phi12}  \\
    $\omega$            &  Angular frequency of the AC signal, s$^{-1}$
\end{tabular}

\newpage

\end{document}